\documentclass[preprint,showpacs,preprintnumbers,amsmath,amssymb]{revtex4}

\usepackage{graphicx}
\usepackage{dcolumn}
\usepackage{bm}

\begin{document}


\title{Growth mechanisms of vapor-born polymer films}

\author{I.J. Lee}
\email{ijlee@chonbuk.ac.kr}
\author{Mira Yun}
\author{Sang-Min Lee}
\author{Ja-Yeon Kim}

\affiliation{ Department of Physics, Research Institute of Physics
and Chemistry, Chonbuk National University, Jeonju, 561-756, Korea}

\date{\today}

\begin{abstract}

The surface morphologies of poly(chloro-p-xylylene) films were
measured using atomic force microscopy and analyzed within the frame
work of the dynamic scaling theory. The evolution of polymer films
grown with fixed experimental parameters showed drastic changes of
dynamic roughening behavior, which involve unusually high growth
exponent ($\beta =$ 0.65$\pm$ 0.03) in the initial growth
regime, followed by a regime characterized by $\beta\sim$ 0, and
finally a crossover to $\beta$ = 0.18$\pm$ 0.02 in a steady growth
regime. Detailed scaling analysis of the surface fluctuation in
Fourier space in terms of power spectral density revealed a gradual
crossover in the global roughness exponent, analogous to a phase
transition between two equilibrium states, from a morphology defined
by $\alpha$=1.36$\pm$0.13 to the other morphology characterized by
$\alpha$=0.93$\pm$0.04 as the film thickness increases. Our experimental results which significant deviate from the well established descriptions of film growth clearly exhibit that the dynamic roughening of polymer film is deeply affected by strong molecular interactions and relaxations of polymer chains.

\end{abstract}
\pacs{68.55.Ac, 68.47.Mn, 81.15.Aa, 68.15.Kk, }
\maketitle

\section{Introduction}

    Organic electronic devices have attracted considerable attentions
on account of their wide range of present and potential
applications, including flexible, transparent, and large-area
electronic systems~\cite{stutzmann,nomura,someya}. Owing to numerous
endeavors to enhance the functionality of the organic devices,
several important aspects of the growth dynamics of organic
semiconductor~\cite{heringdorf,drummy,mayer} and the chemical
effects at the semiconductor/polymer
interface~\cite{pernstich,youn,kim} are now quite well understood.
But relatively little is known about the basic principle governing
the temporal evolution of the growth front of polymer
film itself~\cite{collins,hachen,zhao}. In general, the performance
of the electronic devices based on organic semiconductors is
sensitive to the roughness of dielectric film ~\cite{steudel}. Here,
we report on the surprisingly rich growth dynamics and morphological
characteristics of polymer films grown by vapor deposition. Even
with fixed experimental conditions, the dependence of the
morphological parameters as a function of the film thickness
revealed unprecedented crossovers in the patterns of growth fronts
and in their temporal evolutions. The present findings provide a
spectacular example of manifestation that the instabilities in the
morphological patterns in polymer films were driven by strong
molecular interactions and relaxations.

    Poly(chloro-p-xylylene) (PPX-C) is a transparent plastic film,
and is the most widely used member of polyxylene derivatives (known
by the trade name Parylene) due to its excellent chemical and
physical properties. Parylene is used in a broad range of areas,
including electronics, medical, aerospace, and industrial
applications~\cite{beach1}. In particular, the capacity of PPX-C to
form uniform coating, as well as its high dielectric strength
($>$500V/1$\mu$m)~\cite{beach1} make this polymer film a promising
candidate for use as a dielectric insulator in applications
requiring a thin, transparent, and flexible
film~\cite{nomura,someya}.

\section{Experiment}

    PPX-C were deposited in a custom built chemical vapor deposition
reactor consisting of a sublimation furnace, a pyrolysis furnace,
and a deposition chamber backed by a diffusion pump. Dimmer molecules
(di-chloro-p-xylylene) of granular type were sublimed at 120
$^{\circ}$C and then cracked into monomers in the pyrolysis furnace
at 660 $^{\circ}$C. The monomers were subsequently condensed and
polymerized on SiO$_{2}$/Si substrates in the deposition chamber,
which is held at room temperature. The initial interface width (or
rms roughness) of the clean SiO$_{2}$ surface was typically given as
1.5 ${\AA}$. Several atomically clean substrates were arranged on a
substrate holder which was equipped with a grease-sealed shutter to
prevent impurities and volatile contaminants of the dimmer molecules
(di-chloro-p-xylylene) from condensing on the substrates during the
evacuation of the entire deposition system and during the initial
warming-up of the sublimation chamber. We found that the use of a
substrate shutter providing water leak-tight sealing is crucial to
monitor the early stage of island formation in a steady experimental
condition. The growth of film stops when closing the shutter and
filling the chamber with nitrogen gas to 1 atmospheric pressure.
During the deposition, the pressure was typically in the range of
1-3 mTorr, which gave a growth rate of 20-30 nm/min. Each film grown
with the same experimental condition was used for various
characterizations of the surface, including spectroscopic
ellipsometry for determination of the film thickness and x-ray
photoemission spectroscopy (XPS) for chemical analysis. The surface
morphology was measured using atomic force microscopy (PSIA, model
XE100) in a non-contact mode. Topographic AFM images were taken from
several different locations on each PPX-C film with various scan
sizes in the range of 0.5$\times$0.5 to 5$\times$5 $\mu$m$^2$
depending on the film thickness. X-ray photoelectron spectroscopy
(XPS) measurements (AXIS Nova from Kratos) on several films showed
that they had chemical compositions very close to the formula
composition of PPX-C, without any trace of oxygen contamination. The
lack of oxygen contamination, which occurs due to the termination of
polymer chain-ends by hydroxyl or carbonyl groups during or after
the deposition, indicates that our polymer films were of
exceptionally high quality.

\section{Results and discussion}

\begin{figure}
\includegraphics[width=7.5cm]{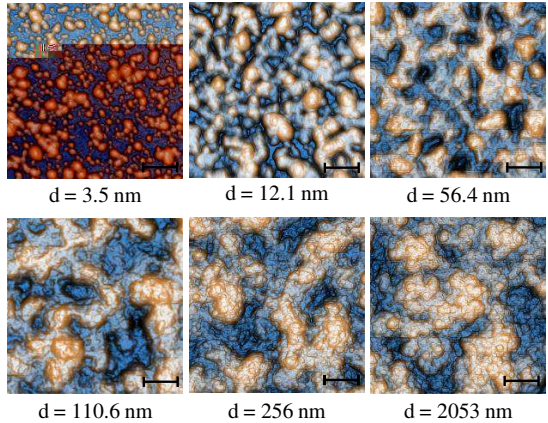}
\caption{Evolution of topographic AFM images of PPX-C films. The
scale bar corresponds to 200 nm in all panels.}\label{fig1:epsart}
\end{figure}

    Representative AFM images taken at various stages of PPX-C film
growth are shown in Fig.1. Islands with rounded edges form during
the early stage of growth and then coalesce as the thickness
increases. Height histograms obtained from the AFM images indicate
three characteristic regimes: 1) the partially covered regime
(thickness, $d <$ 14 nm), in which the bare surface height
contributes to the histogram (see the inset of Fig.2); 2) the regime
of asymmetric height distribution associated with deep valleys (14
nm $< d <$ 150 nm); and 3) the continuous growth regime above 150
nm, where the height distribution is symmetric. For the partially
covered regime ($d <$ 14 nm), the thickness of the film was
determined by averaging the mean-height values from the histograms
of several AFM images taken from different locations on each film
after the contributions from the bare substrate had been subtracted.

    The interface width of the film at different stage of growth is
shown in Fig.2 (a). The data were obtained by measuring the root
mean square fluctuation in height \emph{h},
$W(L,d)=\sqrt{\langle[h(r)-\langle h\rangle]^{2}\rangle}$ where
\emph{L} is the scan size, and $\langle\cdot\cdot\cdot\rangle$
denotes a statistical average over the whole scanned area. Although
the roughening processes that occur during the growth of film are
microscopically diverse and complex in nature, it has been well
established that the evolution of interface width follows a simple
dynamic scaling called Family-Vicsek scaling relation~\cite{family},
expressed as $W(L,d)\sim L^{\alpha}f(d/L^{z})$ where
$z=\alpha/\beta$. Determination of the scaling exponents, namely the
roughness exponent $\alpha$ and the growth exponent $\beta$, allows
one to identify the universality class of the growth process under
study~\cite{barab,krug}. Typically, the evolution of interface width
has two regimes depending on the length scale $L$, that is
$W(L,d)\sim d^{\beta}$ for small film thicknesses and $W(L,d)\sim
constat$ (saturate) for large film thicknesses when the correlation
length $\xi$ is comparable to the system size $L$. Thus, under an
experimental condition where $\xi\ll L$, as in our case (see Fig.
4), the interface width as a function of film thickness is expected
to follow a simple power-law dependence without saturation. In PPX-C
film, however, the interface width as a function of film thickness
(Fig. 2(a)) seems far more complicated than what we would expect
from the typical growth process. Rather, as noticed in the
histograms of the AFM images, three characteristic regimes are
observed as the film grows. In the early rapid growth regime, the
interface width grows with $\beta$=0.65, much faster than the random
deposition limit of stochastic roughening ($\beta$=0.5). Upon
complete coverage of the substrate near $d\sim$10 nm, the film grows
with $\beta\sim 0$. In this regime, the interface width does not
evolve with the film thickness, which is likely due to preferential
filling of deep valleys (the valley filling effect). \emph{Note that
this is a rare display of an extreme case of valley filling effect,
which is strikingly different from the phenomenon described by the usual curvature-driven diffusion model~\cite{amar2,dgk} or the
fourth-order linear MBE equation~\cite{barab,conserve}}. Finally, the film enters the continuous growth regime, in which the interface width
increases steadily with a new power-law of $\beta$=0.18.

\begin{figure}
\includegraphics[width=7.5cm]{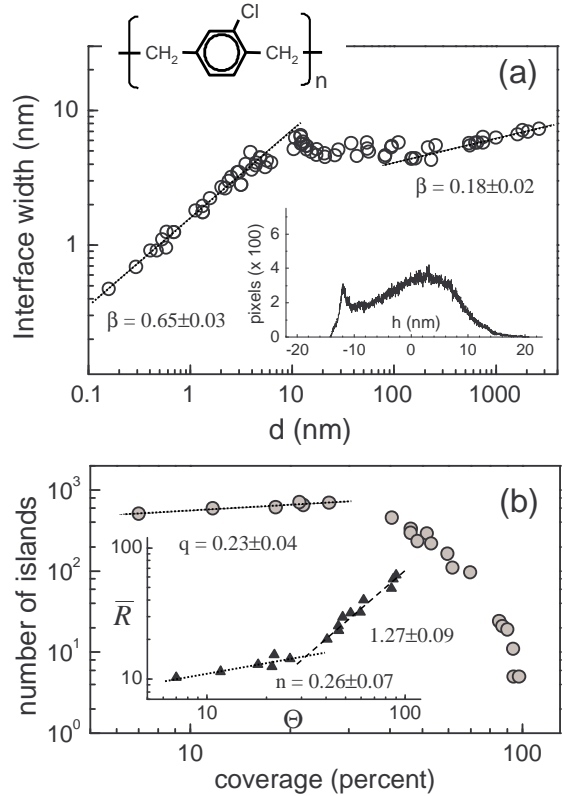}
\caption{Top panel (a), interface width \emph{vs} thickness of the
PPX-C film. The dotted lines are guides for the eyes. Top inset shows
the chemical structure of PPX-C. The bottom inset displays the height
histogram of $d$ = 11.70 nm. Bottom panel (b), number of island as
function of coverage $\Theta$. The inset(filled triangles) shows the
average island size $\bar{R}$ as function of $\Theta$.}
\label{fig2:epsart}
\end{figure}

    Continuum equations describing film growth involve terms for
deposition, desorption, and various diffusions. A crossover in the roughening process can be expected when the competition between different growth mechanisms generates a characteristic length scale in which a particular effect dominates the process~\cite{barab,horo,silva}. With a choice of appropriate coupling strength, these models display crossover effects from one dynamics at small time (or short length scale) to another dynamics at long time (or large length scale). However, the evolution of growth dynamics exhibiting these proposed crossover is hard to realize in practice when experimental conditions are fixed, since the crossover length scale (or deposition time) is usually given as either too short or too long to reach within usual experimental ranges~\cite{barab}. In fact, a few cases of reported crossover effect, such as the laser deposited polymer film~\cite{hachen} and the chemical vapor grown SiO$_{2}$ film~\cite{ojeda}, were attributed not to the competing growth mechanisms but to nonlocal shadowing effects that drive the growth unstable in early time with unusually high growth exponent ($\beta\geq0.5$) before the systems undergo continuous stable growth. Particularly, in the studies of polymer films~\cite{hachen} grown by pulsed laser deposition, the crossover was explained in terms of a transition from a single particle character to a diffusion-driven continuous growth process when the film thickness becomes comparable to the size of the large initial polymer fragments impinging on the flat substrate. The mechanisms underlying the crossovers observed for BPAPC film is likely to be different from that observed for the PPX-C film in the present work because the vapor deposition process of PPX-C does not generate large polymer particles in air. Nevertheless, considering
that film growth is a bottom-up process, we expect the structure and
form of islands on an initially flat surface to exert profound
effects on the patterns and dynamics of the subsequent growth front.
In following, we are trying to convince that the intrinsic nature of
polymerization process~\cite{beach2,fortin} characterized by strong interaction between extremely reactive monomer molecules and free-radical polymer chain-ends, and relaxation of polymer chain through inter-polymer interactions is responsible for the observed unusual growth behavior.

    A detailed analysis of the morphology of the PPX-C films during
the early stage of growth is shown in Fig.2(b). Data were obtained
from 1 $\mu m^{2}$ AFM images comprised of 512 pixels. Island growth
generally involves growth of existing islands, the creation of new
islands, and the coalescence of islands. According to a theoretical
model for molecular beam epitaxy(MBE) based on the point-island rate
equation~\cite{zuo,amar}, the number of island($N$) and the mean
linear island size($\bar{R}$) are expected to depend on the coverage
following the power-law $N(\Theta)\sim \Theta^{q}$ and $\bar{R}\sim
\Theta^{n}$, respectively. When the islands are small and isolated
in the early stage of growth, the number of islands increases as the
coverage increase. During the later stage of growth, when the
coverage approaches a full layer, the number of islands rapidly
decreases due to the merging of clusters. In the intermediate regime
where both island nucleation and coalescence occur simultaneously,
the balance between the two processes determines the values of the
exponents $q$ and $n$ and leads to the relation $q+2n=1$. Fig.2(b)
shows that island coalescence is prominent at coverages above  $~30
\%$, as indicated by the rapid reduction of $N$ and steep increase
in $\bar{R}$. However, the measured exponents below the coalescence
regime are too low to satisfy the simple relation. The main reason
for the failure of the MBE description is that the molecules that
land on top of an island contribute to neither $N$ nor $\bar{R}$.
Moreover, the observation that the mean aspect ratio of islands
($d/\bar{R}$) grows almost linearly with the coverage (not shown)
indicates that the strong over-growth behavior persists throughout.
Because the diffusion barrier (or Schwoebel barrier~\cite{schwo}) is
usually negligible at the step edge of an amorphous polymer film due
to the lack of well defined atomic steps at the surface~\cite{yang},
the strong molecular interactions between strongly reactive monomer
and free-radical polymer chain-ends, and relaxations (or inter-polymer interactions) should play a major role in the unusual over-growth behavior and result in the rapid roughening in our system . Indeed, the rounded edges of the islands (see Fig.1) and the unusual overgrowth behavior, which lacked in the previous Monte Carlo study~\cite{zhao2} considering only monomer surface diffusion, should be a counterexample that proves the presence of strong relaxations which cause entanglement of polymer chains in an way of maximizing the number of nearest neighbors.

    Once the surface is nearly covered, the deep valleys
between islands, where the density of extremely reactive chain-ends
is high, are preferentially filled by monomer radicals. After the
large variation of local curvature of the interface diminishes, the
growth process finds a new equilibrium, which leads to the final
transition to the continuous growth regime. The valley filling
effect generating an opposite effect of shadowing~\cite{karuna,yu}
is indeed another manifestation of the presence of strong molecular
interactions and long diffusion length in polymer film
growth. It is important to note that the nonlocal shadowing effect and step edge barriers, well known causes for the most of rapid roughening ($\beta>0.5$) phenomena~\cite{schwo,yang,karuna,yu} are practically excluded in PPX-C system as a possible explanation for the rapid growth of the interface width ($\beta=0.65$). We also like to point out that the coarse graining mechanism observed in an interesting simulation study by Vree \emph{et al.}~\cite{vree} who considered polymer interactions with the reptation type of relaxation~\cite{degen,rubin} nicely captures an essential part of the observed rapid roughening behavior. Even though, those inter- and intra-polymer interactions produce effects similar to the effects of shadowing and step edge barriers, we expect that the former will cause unusual morphologies and polymer configurations. In following, we deal in detail about the peculiar morphological characters found in polymer films.

\begin{figure}
\includegraphics[width=8.5cm]{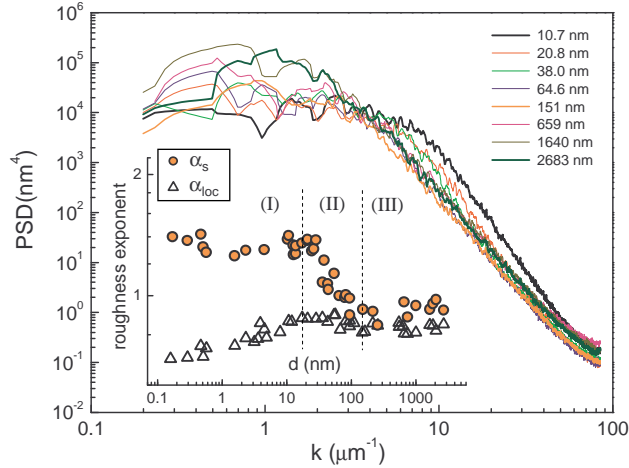}
\caption{The PSD at various film thicknesses. All of the PSD spectra
are based on the AFM images of 3 $\mu m^{2}$ scans. The inset shows
the thickness dependence of the spectral($\alpha_{s}$) and the local
($\alpha_{loc}$) roughness exponents. }\label{fig3:epsart}
\end{figure}

    The unusual dynamics of the roughening process is also reflected
in the evolution of the roughness exponents that characterizes the
surface morphology in the stationary regime. Power spectral density
(PSD) plots recorded at various thicknesses are shown in Fig.3. The
\emph{k}-dependence in the high-\emph{k} regime crosses over to a
saturation (\emph{k}-independent) regime at $k_{c} \sim  1/\xi$ as
the surface features lose their correlation. The value of $k_{c}$
decreases with the film thickness as $k_{c}\sim 1/d^{1/z}$ where $z$
is the dynamic exponent. The best fit of the $k_{c}$ \emph{vs} $d$
plot was achieved using a value of $1/z=0.21\pm0.03$ which was
consistent with the value obtained from the real space relationship
between the correlation length and the thickness $\xi\sim d^{1/z}$
(see Fig. 4)~\cite{corr}. The dynamic scaling behavior of the
two-dimensional surface is modified in the PSD as $S(k) \sim
k^{-(2\alpha_{s}+2)} d^{2(\alpha-\alpha_{s})/z}$ for $k\gg
1/\xi$~\cite{lopez,ramasco}, in which $\alpha_{s}$ and $\alpha$ are
the spectral and the global roughness exponent, respectively. The
spectral roughness exponents, which were obtained
($\alpha_{s}=(m-2)/2$) by measuring the slope ($m$) of the
\emph{k}-dependent PSD plot in the high-$k$ regime, are shown as
filled circles in the inset of Fig.3. The local roughness exponents
marked by open triangles were obtained by fitting the
height-difference correlation function
$H(\textbf{r})=\langle[h(\textbf{r})- h(\textbf{0})]^2\rangle^{1/2}$
to $\rho r^{\alpha_{loc}}$ for small $r$ regime ($r \ll
\xi$)~\cite{barab}. The self-affinity of the interface, a
conventional assumption of the normal scaling behavior, is lost when
the global scaling differs from the scaling behavior of the local
interface width, that is, $\alpha\neq\alpha_{loc}$.

\begin{figure}\includegraphics[width=6.5cm]{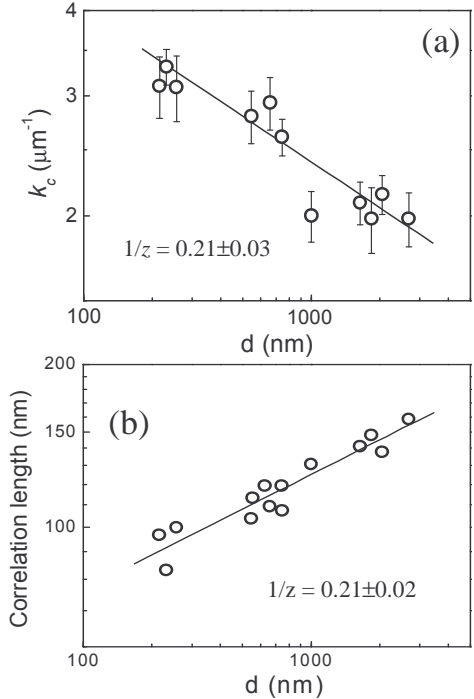}
\caption{(a) The thickness dependence of the crossover $k_{c}$ obtained from PSD functions of the AFM images of 3$\times$3 $\mu$ m$^{2}$ size. The
linear fit (solid line) yields the dynamic exponent 1/z=0.21$\pm$0.03.
(b) The correlation length as a function of film thickness. The slope
of the linear best-fit is displayed as a solid line and yields
1/z=0.21$\pm$0.02.} \label{fig4:epsart}
\end{figure}

In regime (I), the PSD in high-\emph{k} region varies negligibly with the film thickness ($\alpha = \alpha_{s}$) and $\alpha$ is substantially different from $\alpha_{loc}$, which lead to the relation $\alpha_{s}=\alpha\neq\alpha_{loc}$. The scaling behavior of the regime (I) is thus anomalous and belongs to the super-roughing process since
$\alpha=1.36\pm0.13>1$. In the valley filling regime (II), the dynamic
scaling behavior is quite complex that the PSD curves shift downward
($\alpha<\alpha_{s}$) and the slope of the \emph{k}-dependent PSD
continuously decreases with increasing film thickness as shown in
Fig.3. The scaling relationship $\alpha_{s}\neq\alpha\neq\alpha_{loc}$ is satisfied in this intermediate regime and, interestingly, does not belong to any of the classes that have been identified for interfacial growth~\cite{ramasco,lopez,barab}. This unprecedented scaling behavior has not been found in usual atomic deposition process and are likely associated with the intrinsic properties of polymer formation. But the precise role of strong inter- and intra-polymer interactions in the unusual scaling behavior is yet to be unfold. It is also very hard to understand the fact that, in this new class of growth regime, the local exponent remains the same while the global exponent continuously decreases with the film thickness. In the continuous growth regime (III), the PSD curves are thickness invariant, indicating that $\alpha=\alpha_{s}$ and the scaling exponents are obtained as $\alpha_{s}= 0.93\pm0.04$ and $\alpha_{loc}=0.85\pm0.03$. The small difference between $\alpha_{s}$ and $\alpha_{loc}$ appears to be real since multi-affine behavior is clearly visible in this regime~\cite{multi}. In other words, the interface cannot be characterized by a single roughness exponent~\cite{barab}. The dynamic scaling behavior of the regime (III) is similar to that of the regime (I) in which the relation $\alpha_{s}=\alpha\neq\alpha_{loc}$ characterizing the super-rough surface holds. It is interesting to notice that, even though there is a large difference in the global roughness exponent, both the regime (I) and (III) share the same dynamic scaling properties,
which suggest that the strong molecular interaction resulting in
the rapid increase as well as the saturation of the interface width
persists throughout the entire growth process.

    As we may expect from the complex scaling properties and various
crossovers which are significantly different from those of
conventional film growth such as MBE, assigning an universality
class for the polymer film growth is not trivial. The measured exponents $\alpha$=0.93$\pm$0.04, $\beta$=0.18$\pm$0.02, and $1/z$ =0.21$\pm$0.02 in the continuous growth regime compare only marginally, at best, with those obtained from both the linear ($\alpha$=1, $\beta$=0.25, $1/z$ = 0.25)~\cite{amar2,dgk} and the nonlinear MBE equation ($\alpha$=0.67, $\beta$=0.2, $1/z$=0.3)~\cite{ld,dp,wv,dt} in 2+1 dimensions. The clear differences between the observed and the compared are a strong indication that polymer film growth is very different from the conventional MBE growth process. If we neglect all the detailed differences, the local curvature-induced diffusion term ($K\nabla^{4}h$) contained in both the MBE models is responsible for the conventional scaling in reciprocal space ($\alpha=\alpha_{s}$) and the anomalous behavior in real space
($\alpha\neq\alpha_{loc}$)~\cite{dgk,lopez}, as we have observed.
The involvement of curvature-driven diffusion term in polymer
growth seems plausible because local valleys of the interface where
density of active chain-ends or number of nearest neighbors is high
makes a favorable place for both monomer radicals and usual atoms to
stick. However, we note that a series of higher order nonlinear
terms must be included in the picture to account the unusual
multi-scaling behavior~\cite{dp,krug2}. The reasoning is partly consistent with the argument of Punyindu \emph{et al.}~\cite{puny} against the conclusion drawn from previous studies of similar polymer films (PPX) in which nonlocal bulk diffusion was suggested~\cite{zhao} as a governing mechanism in the continuous growth regime.

\section{Summary}

    We have studied the kinetic roughening of vapor-deposited PPX-C
films over a film thickness range encompassing more than three
orders of magnitude. The rapid growth in the early stage and crossover to the saturation of interface width which significantly deviate from the conventional MBE description were understood in line with strong molecular interaction and relaxation of polymer chains. The characteristic exponents in the continuous growth regime were crosschecked with independent methods and consistently obtained as $\alpha$=0.93$\pm$0.04, $\beta$=0.18$\pm$0.02, and $1/z$ = 0.21$\pm$0.02. As we have discussed, the scaling exponents cannot be directly related to any known universality classes based on conventional atomic deposition. In light of previous theoretical studies of polymer growth~\cite{zhao2,vree} which specifically designed for a linear chain polymer, our experimental results suggests that a successful polymer model should incorporate the intrinsic properties of polymer growth such as strong interaction between monomer radicals and polymer chain-ends, and relaxation of polymer chains. Certainly, polymer systems should provide a myriad of possibilities for learning more about the formation of rough fronts and their evolution beyond the MBE process. Finally, since the peculiar growth behaviors of PPX-C films come from the intrinsic properties of polymerization process, we expect to see unusual growth behavior in other linear chain polymers. However, we note that different chemical compositions of polymer should provide ranges of interaction strength which essentially determines the behavior of kinetic roughening (see, for example, ref~\cite{zhao}, in which unusual growth behavior was argued for bulk diffusion). So far, our studies have raised more questions than answers regarding the growth mechanism of polymer film. We hope the present experimental study will stimulate the interest in polymer growth.

\begin{acknowledgments}
We thank S.M Jeong and M.S Ha for valuable discussions. This work was
supported by the Korea Science and Engineering Foundation Grant No.
R01-2006-000-10800-0 and by the Korea Research Foundation
(KRF-2005-015-J07501).
\end{acknowledgments}


\begin{thebibliography} {}
\bibitem{conserve} The curvature diffusion models with conserved noise which possibly give the saturated interface width in 2+1 dimensions are not relevant in our case because monomers are permantely attached to polymer chain-ends through chemical bonding.
\bibitem{corr} The correlation length can be accurately determined from
the relation $C(\xi)/C(0)=e^{-1}$ after evaluation of the two dimensional
correlation function $C(\mathbf{r})=\langle h(\mathbf{r})h(0)\rangle$ for
each of the AFM images.
\end{thebibliography}
\end{document}